# Vibration Mitigation in Partially Liquid-Filled Vessel using Passive Energy Absorbers

*Maor Farid[1], Nissim Levy[2] and Oleg Gendelman[1,*]*

[1] *Faculty of Mechanical Engineering, Technion – Israel Institute of Technology*

[2] *Nuclear Research Center of the Negev (NRCN)*

\* - contacting author, e-mail:ovgend@tx.technion.ac.il

## Abstract

This paper treats possible solutions for vibration mitigation in reduced-order model of partially-filled liquid tank under impulsive forcing. Such excitation may lead to hydraulic impacts applied on the tank inner walls. Finite stiffness of the tank walls is taken into account. We explore both linear (Tuned Mass Damper) and nonlinear (Nonlinear Energy Sink) passive vibration absorbers; mitigation performances are examined numerically. The liquid sloshing mass is modeled by equivalent mass-spring-dashpot system, which can both perform small-amplitude linear oscillations and impact the vessel walls. We use parameters of the equivalent mass-spring-dashpot system for well-explored case of cylindrical tanks. The hydraulic impacts are modeled by high-power potential and dissipation functions. Critical location in the tank structure is determined and expression of the corresponding local mechanical stress is derived. We usefinite-elemet approach to assess  the natural frequencies for specific system parameters and to figure out possibilities for internal resonances. Numerical evaluation criteria are suggested to determine the energy absorption performance.

# 1. Introduction

Vessels filled with liquid are used in many fields of engineering, including nuclear [1], vehicle[2,3] and aerospace industries[4], for storage of chemicals, gasoline, water, and different hazardous liquids[5]. External excitations may cause well-known dynamical effect of liquid sloshing. This effect can take place in liquid cargo on highways, cruises or in stationary vessels exposed to earthquake. Dynamic loads related to the liquid sloshing may have direct and rather strong hazardous effect on the vessel stability and robustness. Many methods of seismic analysis of tanks are currently used by researchers and have been adopted by a number of industry standards and guides such as ACI 350.3-06 and ACI 371R-08 covert seismic design, which are based on the simplified methods evolved from earlier analytical work by Housner 1960, A.S. Veletsos 1977-1984, M.A. Haroun 1981 and Shivakuinar 1997, and others. Of these, the best known is Housner's pioneering work, published in the early 1960s by the Atomic Energy Commission. Housner's method was adopted by many codes in the world and by a number of industry standards and has served as a guideline for most seismic designs of liquid storage tanks. According to Housner's theory, which is a simplified method of analysis, the tank deformation is negligible with respect to the liquid motion. Hence, the analysis is taking into account only the liquid motion relative to the tank and assuming that the tank motion is proportional to the ground excitation. Housner observed that in certain parts of the tank structure, the sloshing of the water was the dominant factor, whereas for other parts, the sloshing had a small effect.

So far, detailed analytical solutions are limited to small-amplitude sloshing in rectangular and cylindrical vessels. While being most interesting and potentially hazardous, high-amplitude liquid sloshing in cylindrical tanks still lacks complete analytic description. The reason is that the liquid in the tank is continuous system with infinite number of degrees of freedom, and its boundary conditions on the free surface are nonlinear and time-dependent. Nevertheless, loads created by high-amplitude liquid sloshing are so crucial for designing the containing vessel, its supports and payload limitations [6], that a number of approximate phenomenological models were developed in order to get at least qualitative insight into this phenomenon.

In well-known phenomenological models, the sloshing dynamics in a partially-filled liquid tank is modeled by a mass-spring-dashpot system or a pendulum, when each sloshing mode is modeled by a different modal mass. One can easily understand that the former dynamics is less complex, since it involves only one dimensional dynamics and interactions, however it fails to represent vertical liquid motion (e.g. water jets[7]) and vertical excitation, that are better represented by the pendulum model (parametric excitation of liquid-filled vessel modeled by high-exponent potential pendulum by El-Sayad[8], Pilipchuk and Ibrahim[9]). Moreover, mass-spring-dashpot system is more common among engineering design regulations and analyzes, as shown by Malhotra et al. [10]. It is worth mentioning that parameter values for both models mentioned above are presented by Dodge [11] and Abramson [12].

This study involves seismic-induced horizontal excitation. Therfore, we take into account only horizontal internal forces. Consequently, the sloshing dynamics in a partially-filled liquid tank with total mass M is modeled by a mass-spring-dashpot system with mass $m$, stiffness $k$, linear viscosity of $c$ and a transversal coordinate $y$ with respect to the vessel centerline. In this simplified model, three dynamic regimes can take place:

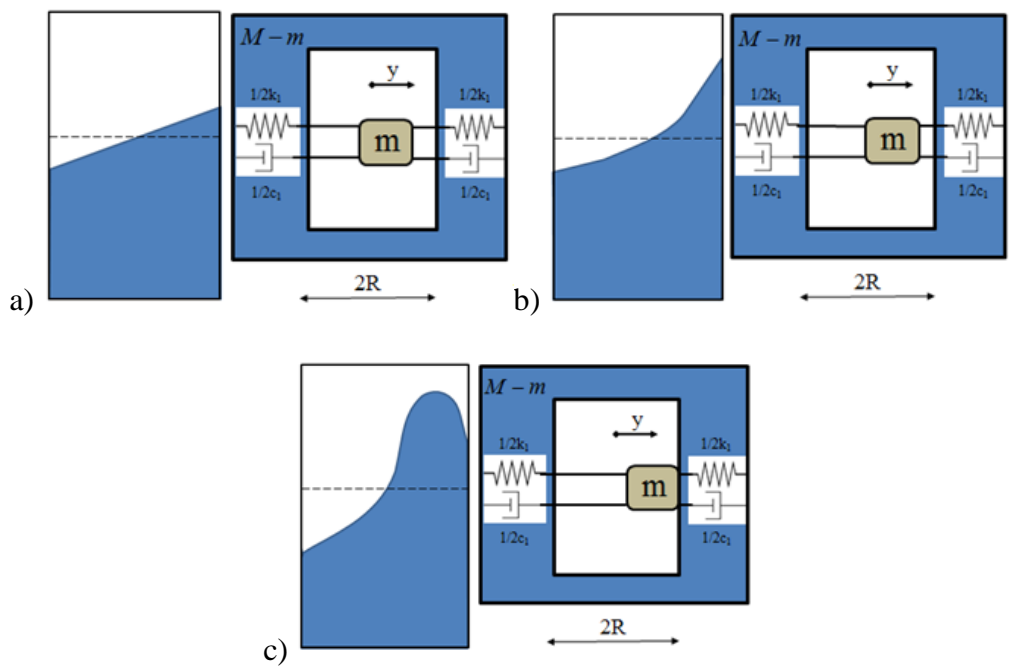

*Figure 1- Regimes of liquid free-surface motion and their equivalent mechanical models.*

(a) The liquid free surface performs small oscillations around its trivial stable equilibrium and remains planar. This regime can be successfully described by a linear mass-spring-dashpot system performing small oscillations.

(b) Relatively large oscillations in which the liquid free surface does not remain planar. This motion is described by a differential equation with weak nonlinearity, and can be treated by perturbation methods [9,13,14]. In this regime the equivalent mass-spring-dashpot system is considered to perform moderate oscillations, so that a cubic stiffness spring addition is reasonable, and the nonlinearity can be treated as weak.

(c) The free liquid surface is urged into a strongly nonlinear motion, related to liquid sloshing impacts with the tank walls. This regime can be described with the help of a mass-spring-dashpot system, which impacts the tank walls.

High-amplitude sloshing can cause hydraulic jumps. In this case (Figure 1(c)) major hydraulic impacts can act on the vessel structure walls [15]. Despite obvious practical interest, methods for evaluation of the impact in this case are not well developed, and rely primarily on data of direct experiments [12]. Hydraulic jumps and wave collisions with vessel shell are the source of strong non-linearities in the system, since collisions with rigid or elastic masses during vibration cause fast velocity changes. Hence, even if the sloshing appears due to simple harmonic excitations, the response may be neither harmonic nor periodic. Authors of paper [16] suggest application of methods developed for analysis of vibro-impact motion. In this work another approach is applied. We use high order smooth potential function [17,18] to model the interaction between the free-surface wave and the vessel walls, following Pilipchuk and Ibrahim [9]. The vibro-impact problem of a mass oscillating in a rigid container was solved in previous works by Buzhinskii and Stolbetsov [19] and by Shaw and Shaw[20].

Modeling of free-surface oscillations in rectangular tanks with the help of equivalent pendulum was developed by Graham in 1951 [21]. Equivalent moment of inertia of a liquid in cylindrical containers has been estimated numerically by Partom ([22] and [23]) and verified experimentally by Werner and Coldwell) [24]. Parameters of equivalent mechanical model, which corresponds to the first asymmetric sloshing mode of cylindrical and rectangular tanks were studied by Dodge [11] and Abramson [12]. The nonlinear interaction of the sloshing liquid with elastic tank in conditions of parametric excitation was studied by Ibrahim[25], Ibrahim and Barr[26,27], Ibrahim, Gau and Soundararajan[28], and El-Sayad, Hanna and Ibrahim [8]. Observations and experiments show that in the vicinity of internal resonances violent response might take place. Using a continuous liquid model, similar interactions resulting from horizontal and combined ground excitation were studied by Ibrahim and Li[29], and for parametric excitation by Soundarajan and Ibrahim[30]. Both harmonic parametric excitation and saw-tooth approximation for the horizontal external excitation were applied by Pilipchuk and Ibrahim [9], using a discrete sloshing model.

Liquid sloshing dynamics in partially filled vessels depends on liquid depth. For shallow free-surface, waves and hydraulic jumps show in vicinity of resonance and apply high loads on vessel walls [15]. For high free-surface, hydraulic loads may be applied on the vessel top.

eismic damage modes in cylindrical storage tanks were studied thoroughly by many researchers, e.g. Suzuki[31]. Most of them are related to the tank foundations and side wall, for example: side walls buckling, base plate cracking, anchor bolts failure, local base fracture etc. Tall slim tanks are exposed to bending buckling, causing elephant-foot bulges, when short thick tanks are exposed to shear bucking, related to diamond buckling near the tank base. Those failure modes are explained extensively by Maekawa[6].

Cylindrical shell vibration modes are characterized by longitudinal half wave number $m$ and circumferential wave number $n$. Beam-type modes correspond to the modes with $n=1$ and $m \geq 1$, while the oval-type vibration modes correspond to $n \geq 2$ and $m \geq 1$. Typical vibration modes of a partially-filled liquid vessel are illustrated in Figure 2.

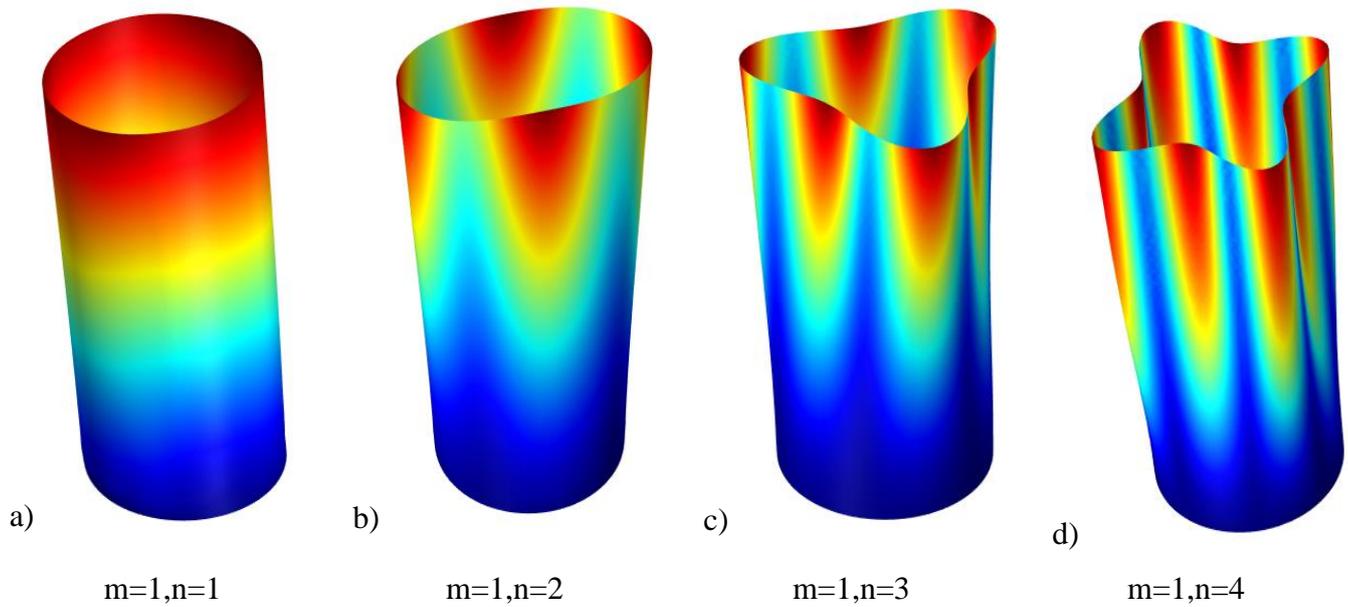

|  a) | b) | c) | d) |
| m=1,n=1 | m=1,n=2 | m=1,n=3 | m=1,n=4 |

*Figure 2 - Typical vibration modes of thin cylindrical shell: m, axial half-wave number; n, circumferential wave number; a) the fundamental axisymmetric beam-type vibration mode (1,1); b)-d) petal-wave form vibration modes.*

According to linear shell analysis, perfectly axisymmetric cylindrical shell vibration modes with $n \geq 2$ are not excited. Moreover, Technical Codes for a Seismic design of Nuclear Power plants (Japan Electric Association [JEA], 2008), the High-Pressure Gas Safety Institute of Japan [KHK], and the Architectural Institute of Japan [AIJ] merely refer to the tank bending vibration modes since they mainly affect the seismic resistance of the cylindrical storage tank, contrary to the oval-type vibration modes. Moreoverthe bending modes are characterized by the motion in the same direction as the sloshing oscillations. Hence, mutual energy exchanges between the vessel and liquid (resonance)are likely. In the subsequent analysis we take into account only the lowest beam-type mode, i.e. the fundamental (1,1) vibration mode (Figure 3).

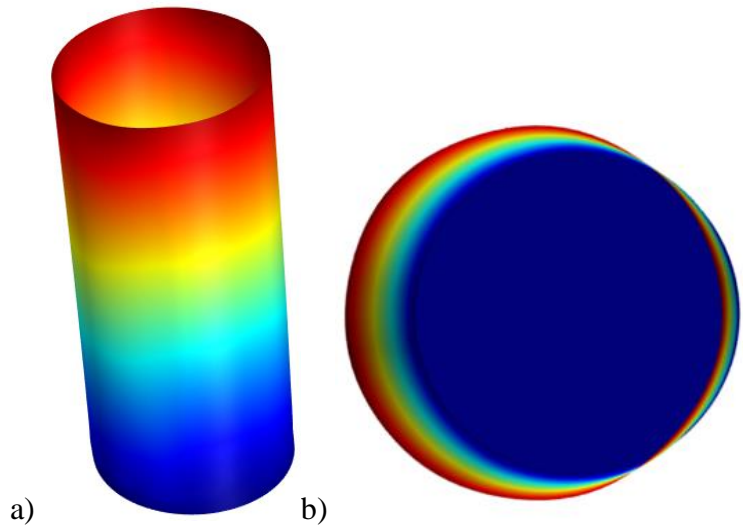

a)      b)

*Figure 3- The fundamental beam-type vibration mode (1,1); a)isometric view, b) top view.*

Reduced order models of the liquid sloshing in cylindrical and rectangular tanks were suggested by Dodge [11]. These models include infinite number of modal masses, corresponding to infinite set of the sloshing modes. Dodge shows that themodal mass rapidly decreases with increasing mode number. Then, one can consider only the asymmetric sloshing modes, which are of major concern since they are associated to horizontal oscillations of the vessel center of mass. Symmetrical modes about the vertical axis are possible, but seem less critical [32]. In the case of lateral excitation, the first asymmetric sloshing mode (1,1) is predominant[33] and is related to the lowest natural sloshing frequency. Consequently, it is reasonable to take into account only the first sloshing mode in the mechanical equivalent model, as long as the excitation frequency is far from the natural frequencies of the higher modes.

Many physical systems are exposed to environmental disturbances. These unwanted phenomena cause undesirable vibration, which one would prefer to suppress.. One solution to this problem is a passive energy absorber (PEA).

Passive energy absorber is a relatively small mass attached to the primary system. The linear PEA is better known as Tuned mass damper (TMD) [34]. When tuned to the primary system's natural frequency, the TMD performs large amplitude oscillations instead of the primary mass, which oscillates with relatively small amplitude. Many well-established engineering designs of vibration absorbers rely on the concept of TMD. However, the TMD dynamics arises essential problems. First, the TMD is a narrowband device, and demands high precision in tuning its mechanical properties; this, in turn, restricts the robustness. Secondly, a technical implicational problem arises due to the large amplitude of the TMD oscillations. Hence, the TMD demands a large housing next to the primary system, which often is not possible, especially in small scale devices. To overcome these restrictions, nonlinear PEA solutions are considered. Nonlinear energy sink (NES) is defined as an essentially nonlinear oscillator with relatively small mass, attached to a primary mechanical system in order to passively absorb the energy of oscillations under various forcing conditions [35–37].The NES can be considered as a nonlinear generalization of the linear vibration absorber, the TMD. A comparison between the TMD and cubic NES performances has been demonstrated by Vakakis et al. [38] for linear oscillator under impulsive forcing.

In this paper, in section 1 we describe TMD and NES system coupled to the equivalent modelof the partially-filled cylindrical storage tank. The parameter values are calculated for a 75% full aluminum water tank. The equations of motion for a generic ground excitation and a resulting expressions for the critical stresses are formulated. Conditions for internal resonances are shown numerically in section 2. In section 3 we use Finite-Element (FE) analysis to calculate the tank-liquid system natural frequencies and compare these results with literature values. We exam both TMD and NES performance and their contribution to the tank seismic mitigation. In section 4 PEA numerical optimization evaluation criteria and objective function are chosen. Numerical optimization is shown in section 5 and the TMD and NES most efficient dynamic regime are revealedand analyzed.

## 2. Model description

### 2.1 Equations of motion

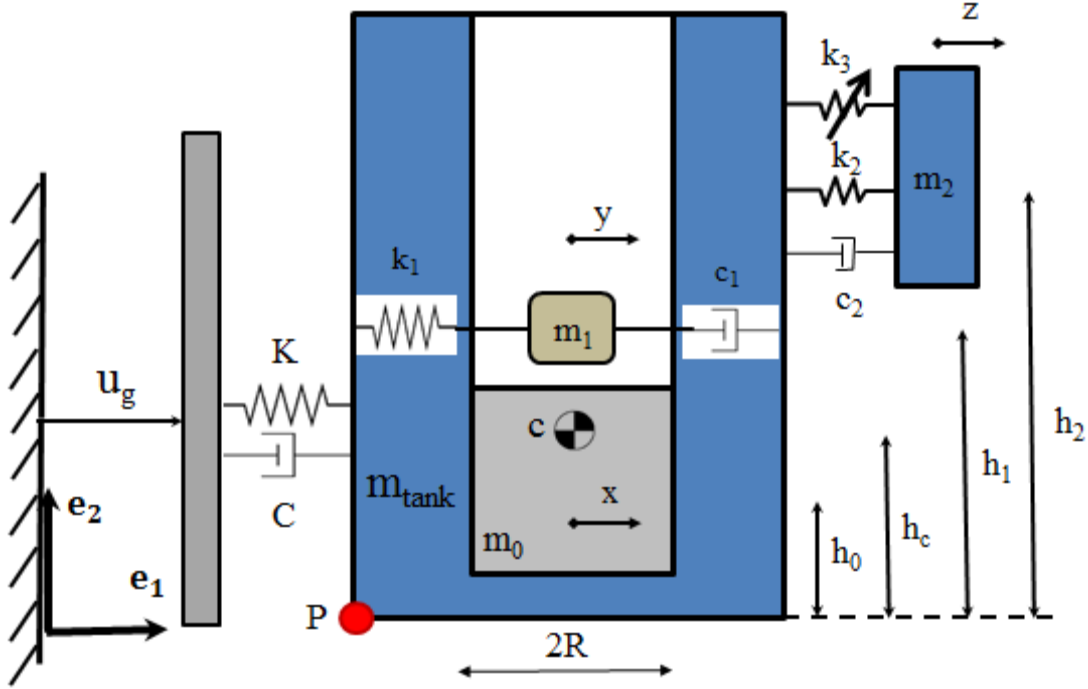

*Figure 4- Scheme of multiple sloshing modes in cylindrical tank represented by mass-spring-dashpot system interacting with structure walls.*

In the following Section, we introduce the equivalent mechanical model for liquid sloshing in cylindrical tank with height, radius and wall thickness of $H$, $R$ and $t_w$, shown in Figure 4. Where $u_g = A\cos(\omega_{ext} t)$ is the horizontal ground excitation applied on the vessel base with respect to inertial frame of reference $\mathbf{e}_1, \mathbf{e}_2$. K and C are modal stiffness and damping of the vessel fundamental (1,1) beam-type mode, respectively. Masses $m_{tank}$ and $m_0$ are the masses of the tank shell and the convective portion of the liquid respectively. $m_1$, $k_1$ and $c_1$ are the mass, linear stiffness and viscous damping coefficients of the asymmetric fundamental sloshing mode[6].

Due to maximal bending moment and local geometric variation in the tank foundation that leads to stress concentration, the critical maximum-stress point P is located on the outer region of the tank base.

The combined tank and 'convective' liquid portion mass, denoted by *M*, is given by the following expression:

$$M = m_0 + m_{\tan k} \qquad (1)$$

Point $c$ is the overall system center of gravity. The combined mass, sloshing-mass and PEA dimensional displacements are denoted by coordinates $x$, $y$ and $z$, respectively. The sloshing mass is allowed to move in a straight cavity along the vessel diameter. The PEA mass $m_2$ is coupled to the tank with linear and cubic springs, with stiffness coefficients of $k_2$ and $k_3$ respectively, and linear viscous damping coefficient $c_2$. The case of the TMD attachment corresponds to $k_2 \neq 0, k_3 = 0$, and the cubic NES attachment - to $k_2 = 0, k_3 \neq 0$.

The analytical expressions for the fundamental asymmetric sloshing mode mass and stiffness $m_1, k_1$ for cylindrical tank, are calculated by Dodge [11] as follows:

$$m_1 = m_F \frac{R}{2.2h} \tanh(1.84 h/R), \quad k_1 = m_F \frac{g}{1.19h} \tanh^2(1.84 h/R) \qquad (2)$$

where $h$ is the liquid depth of the undisturbed liquid free-surface, $m_F$ is the total liquid mass, and $g$ is the gravity coefficient. For cylindrical tank and water density of $\rho_F = 997 \, kg/m^3$, the total liquid mass is expressed as follows:

$$m_F = \pi R^2 h \rho_F \qquad (3)$$

The sloshing dynamics combines infinite number of the sloshing modes. The modal masses of the asymmetric sloshing modes in cylindrical vessels were well documented by Abramson[12]. Figure 5 presents the dependence of the 'static' fluid portion ratio $m_0/m_F$, and modal sloshing mass ratios of the first three sloshing modes versus the depth-radius ratio $h/R$. One can see that all modal masses decrease rapidly with fluid depth exceeding the first. However, as more slender and/or full the vessel, the less significant the first mode modal mass ratio. Consequently, when the base excitation frequency content is far from the higher modal frequencies (2[nd], 3[rd] and so forth), we can consider onlythe frequency of the first sloshing modal mass in the reduced order model.

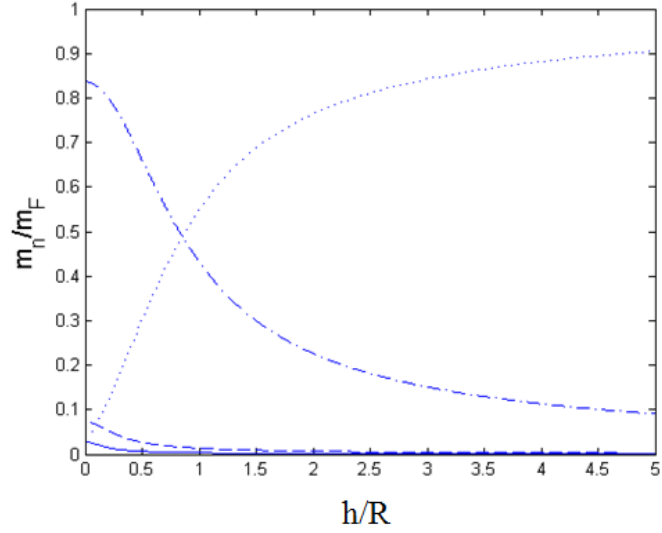

*Figure 5-Ratios of the first three asymmetric sloshing modal masses $m_1$, $m_2$ and $m_3$ and fixed mass $m_0$ to the total fluid mass $m_F$ for cylindrical vessel; dotted-lime: $m_0/m_F$, dashed-dotted-line: $m_1/m_F$, dashed-line: $m_2/m_F$ solid-line: $m_3/m_F$.*

The total fluid mass is expressed as follows:

$$m_F = m_0 + \sum_{n=1}^{\infty} m_n \quad (4)$$

As it was mentioned above, we restrict ourselves by the first sloshing mode; therefore, the expression for the mass of liquid is treated in the following form:

$$m_F = m_0 + m_1 \quad (5)$$

The convective liquid portion mass is given by equations(2), (3) and (5). Following Dodge [11], the sloshing mass and the convective liquid portion mass height with respect to the tank foundations are as follows, respectively:

$$h_1 = \frac{h}{2} - 1.087 R \tanh(0.92 h/R), \quad h_0 = \frac{m_1}{m_0} h_1 \quad (6)$$

Since the empty tank center of gravity is $H/2$, the combined tank-static liquid portion center of gravity height issituated at the height:

$$h_c = \frac{m_{\tan k} H/2 + m_0 h_0}{M} \quad (7)$$

The TMD height $h_2$ is determined by the designer. The system Lagrangian is written as follows:

$$L = \frac{M}{2}(u_{g,t} + x_t)^2 + \frac{m_1}{2}(u_{g,t} + y_t)^2 + \frac{m_2}{2}(u_{g,t} + z_t)^2 - \\ -\frac{K}{2}x^2 - \frac{k_1}{2}(y-x)^2 - \frac{m_1 \tilde{k}}{2}\left(\frac{y-x}{R}\right)^{4n+2} - \frac{k_2}{2}(z-x)^2 - \frac{k_3}{4}(z-x)^4 \quad (8)$$

The three first terms in equation (8) correspond to the tank-'static'-liquid-portion, sloshing mass and PEAs kinetic energy, respectively. The following two term correspond to the linear and nonlinear interactions between the sloshing liquid portion and the tank walls, respectively. The last term correspond to the interaction between the tank and the PEAs.

The dissipation function is as follows:

$$D = \frac{C}{2}x_t^2 + \frac{c_1}{2}(y_t - x_t)^2 + \frac{m_1 \tilde{c}}{2}(2n+1)(y_t - x_t)^2 \left(\frac{y-x}{R}\right)^{2n} + \frac{c_2}{2}(z_t - x_t)^2 \quad (9)$$

The first term in equation(9) corresponds to the dissipation involved in the tank center of mass motion. The following two term correspond to the sloshing liquid relative motion with respect to the tank centerline, and the last term to the PEAs.

For arbitrary external excitation $u_g(t)$, the dimensional equations of motion are as follows:

$$M(u_{g,tt}(t) + x_{tt}) + Kx - \frac{m_1 \tilde{k}(2n+1)}{R}\left(\frac{y-x}{R}\right)^{4n+1} - k_2(z-x) - \\ -k_3(z-x)^3 - k_1(y-x) + Cx_t - c_1(y_t - x_t) - c_2(z_t - x_t) - m_1\tilde{c}(2n+1)(y_t - x_t)\left(\frac{y-x}{R}\right)^{2n} = 0 \\ m_1(u_{g,tt}(t) + y_{tt}) + \frac{m_1 \tilde{k}(2n+1)}{R}\left(\frac{y-x}{R}\right)^{4n+1} + k_1(y-x) + c_1(y_t - x_t) + m_1\tilde{c}(2n+1)(y_t - x_t)\left(\frac{y-x}{R}\right)^{2n} = 0 \\ m_2(u_{g,tt}(t) + z_{tt}) + k_2(z-x) + k_3(z-x)^3 + c_2(z_t - x_t) = 0 \quad (10)$$

We introduce the following non-dimensional time variable: $t_N = \Omega t$, where $\Omega^2 = K/M$ is the vessel natural frequency in units of $rad/\sec$. We rescale the equations by the length $R$ ($\bar{x} = x/R, \bar{y} = y/R, \bar{z} = z/R$). The normalized equations of motion are as follows:

$$\bar{x}'' + 2\bar{Z}\bar{x}' + \bar{x} - \varepsilon_1\beta_1^2(\bar{y}-\bar{x}) - 2\varepsilon_1\beta_1\bar{\zeta}_1(\bar{y}'-\bar{x}') - \varepsilon_2\beta_2^2(\bar{z}-\bar{x}) - \varepsilon_2\bar{k}_2(\bar{z}-\bar{x})^3 -$$
$$-2\varepsilon_2\beta_2\bar{\zeta}_2(\bar{z}'-\bar{x}') - \varepsilon_1\bar{k}(\bar{y}-\bar{x})^{4n+1} - \varepsilon_1\bar{c}(\bar{y}'-\bar{x}')(\bar{y}-\bar{x})^{2n} = -\frac{1}{R\Omega^2}u_{g,tt}(t) \quad (11)$$
$$\bar{y}'' + \beta_1^2(\bar{y}-\bar{x}) + 2\beta_1\bar{\zeta}_1(\bar{y}'-\bar{x}') + \bar{k}(\bar{y}-\bar{x})^{4n+1} + \bar{c}(\bar{y}'-\bar{x}')(\bar{y}-\bar{x})^{2n} = -\frac{1}{R\Omega^2}u_{g,tt}(t)$$
$$\bar{z}'' + \beta_2^2(\bar{z}-\bar{x}) + \bar{k}_2(\bar{z}-\bar{x})^3 + 2\beta_2\bar{\zeta}_2(\bar{z}'-\bar{x}') = -\frac{1}{R\Omega^2}u_{g,tt}(t)$$

where tag denotes a derivation with respect to time scale $t_N$. The non-dimensional parameters governing the system dynamics are as follows:

$$\Omega^2 = \frac{K}{M}, \omega_1^2 = \frac{k_1}{m_1}, \omega_2^2 = \frac{k_2}{m_2}$$
$$\beta_1 = \frac{\omega_1}{\Omega}, \beta_2 = \frac{\omega_2}{\Omega}, \bar{k}_2 = \frac{k_3 R^2}{m_2 \Omega^2}, \varepsilon_1 = \frac{m_1}{M}, \varepsilon_2 = \frac{m_2}{M}$$
$$\bar{Z} = \frac{C}{2M\Omega}, \bar{\zeta}_1 = \frac{c_1}{2m_1\omega_1}, \bar{\zeta}_2 = \frac{c_2}{2m_2\omega_2}, \bar{\xi}_2 = \frac{c_2}{m_2\Omega} \quad (12)$$
$$\bar{k} = \frac{\tilde{k}(2n+1)}{\Omega^2 R^2}, \bar{c} = \frac{\tilde{c}(2n+1)}{\Omega}$$

The TMD non-dimensional parameters $\beta_2$ and $\bar{\zeta}_2$, the NES non-dimensional parameters $\bar{k}_2$ and $\bar{\xi}_2$, and the PEA mass ratio $\varepsilon_2$ are chosen by the designer. Critical damping case corresponds to $\bar{\zeta}_2 = 1$. To simplify the equations of motion, we introduce the following transformations:

$$u = \bar{x} + \varepsilon_1\bar{y} \quad \bar{x} = (u + \varepsilon_1 v)/(1+\varepsilon_1)$$
$$v = \bar{x} - \bar{y} \quad \bar{y} = (u-v)/(1+\varepsilon_1) \quad (13)$$
$$w = \bar{z} - \bar{x} \quad \bar{z} = (u + \varepsilon_1 v)/(1+\varepsilon_1) + w$$

We also rescale the equations with respect to a new time scale: $\tau = t_N/\sqrt{1+\varepsilon_1}$, so in further expressions, dot denotes a derivation with respect to time scale $\tau$.

The transformed equations of motion are as follows:

$$\begin{aligned}
&\ddot{u}+u+\varepsilon_1 v+Z\dot{u}+\varepsilon_1 Z\dot{v}-\varepsilon_2(1+\varepsilon_1)\beta_2^2 w-\varepsilon_2\kappa_2 w^3- \\
&-2\varepsilon_2\beta_2\zeta_2\dot{w}=-\frac{(1+\varepsilon_1)^2}{R\Omega^2}u_{g,tt}(t) \\
&\ddot{v}+u+\left(\varepsilon_1+(1+\varepsilon_1)^2\beta_1^2\right)v+Z\dot{u}+\left(\varepsilon_1 Z+2(1+\varepsilon_1)\beta_1\zeta_1\right)\dot{v}- \\
&-\varepsilon_2(1+\varepsilon_1)\beta_2^2 w-\varepsilon_2\kappa_2 w^3-2\varepsilon_2\beta_2\zeta_2\dot{w}+(1+\varepsilon_1)\kappa v^{4n+1}+(1+\varepsilon_1)\lambda\dot{v}v^{2n}=0 \\
&\ddot{w}-u-\varepsilon_1\left(1+(1+\varepsilon_1)\beta_1^2\right)v+(1+\varepsilon_1)(1+\varepsilon_2)\beta_2^2 w+(1+\varepsilon_2)\kappa_2 w^3- \\
&-Z\dot{u}-\varepsilon_1(Z+2\beta_1\zeta_1)\dot{v}+2(1+\varepsilon_2)\beta_2\zeta_2\dot{w}-\varepsilon_1\kappa v^{4n+1}-\varepsilon_1\lambda\dot{v}v^{2n}=0
\end{aligned}$$
(14)

Where the additional non-dimensional parameters are as follows:

$$Z=\frac{2\bar{Z}}{\sqrt{1+\varepsilon_1}},\zeta_1=\sqrt{1+\varepsilon_1}\bar{\zeta}_1,\zeta_2=\sqrt{1+\varepsilon_1}\bar{\zeta}_2,\xi_2=\sqrt{1+\varepsilon_1}\bar{\xi}_2$$
$$\kappa_2=\bar{k}_2(1+\varepsilon_1),\kappa=\bar{k}(1+\varepsilon_1),\lambda=\bar{c}\sqrt{1+\varepsilon_1}$$
(15)

For a simple, linear, two DOFs case corresponding to TMD attachment only ($\kappa_2=0$) that can be solved and analyzed analytically, the TMD optimal damping coefficient $\bar{\zeta}_2$ for harmonic excitation can bechosen bt analytic method, as shown by Den Hartog [34]. However, in our nonlinear, complex, three DOFs case, analytical approach is no longer possible, and consequently the pseudo-optimal value of $\bar{\zeta}_2$ should be chosen with the help of numerical optimization.

We linearize equation(14) that corresponds to the TMD case ($\kappa_2=0$) with respect to the trivial equilibrium, to yield the following equation in matrix form:

$$\mathbf{I}\ddot{\mathbf{q}}+\mathbf{K}\mathbf{q}=\mathbf{0}$$
(16)

where $\mathbf{q}=(u\ v\ w)^T$, $\mathbf{I}$ is the identity matrix, and $\mathbf{K}$ is the following stiffness matrix:

$$\mathbf{K}=\begin{pmatrix} 1 & \varepsilon_1 & -\varepsilon_2(1+\varepsilon_1)\beta_2^2 \\ 1 & \varepsilon_1+(1+\varepsilon_1)^2\beta_1^2 & -\varepsilon_2(1+\varepsilon_1)\beta_2^2 \\ -1 & -\varepsilon_1\left(1+(1+\varepsilon_1)\beta_1^2\right) & (1+\varepsilon_1)(1+\varepsilon_2)\beta_2^2 \end{pmatrix}$$
(17)

## 2.2. Stresses assessment in critical section

The horizontal forces exerted on the tank walls: $F_0, F_1, F_2$ and their corresponding locations of action with respect to point P: $\mathbf{r}_{c,P}, \mathbf{r}_{1,P}$ and $\mathbf{r}_{2,P}$ are given in appendix A. Since free force diagram is applied on the tank structure- $F_0$ is considered as an external force, while $F_1$ and $F_2$ are internal forces between the tank structure and the sloshing mass and the TMD, respectively. The total shearing force applied on the tank walls is given as follows:

$$F_s \equiv M\left(x_{tt} + u_{g,tt}\right) = F_0 + F_1 + F_2 \tag{18}$$

Since the tank wall area section is $A_S = 2\pi R t_w$, after first time normalization according to $t_N = \Omega t$, the shear stress applied on the tank shell is as follows:

$$\tau_{shear} = F_s / A_s = \frac{M\Omega^2}{2\pi t_w}\left(\bar{x}'' + \bar{u}_g''\right) \tag{19}$$

The system overall center of gravity $cg$ for N concentrated masses system is given by the following formula:

$$\mathbf{r}_{cg} = \frac{\sum_{i=1}^{N} m_i \mathbf{r}_{cgi}}{\sum_{i=1}^{N} m_i} \tag{20}$$

where $r_{cgi}$ is the center of gravity of the $i^{th}$ mass. The overall tank-liquid-TMD system center of gravity is $\mathbf{r}_{cg} = x_{cg}\mathbf{e_1} + y_{cg}\mathbf{e_2}$, where components $x_{cg}$ and $y_{cg}$ are given in appendix A. Parameters $h_c$, $h_1$ and $h_2$ are the heights of masses $M, m_1$ and $m_2$, respectively. The total bending moment applied on moving point P is given by the following equation:

$$\mathbf{M_P} = \mathbf{H}_{P,t} + \left(M + m_1 + m_2\right)\mathbf{r}_{cg,P} \times \mathbf{r}_{P,tt} \tag{21}$$

where $\mathbf{H_P}$ is the system total angular momentum with respect to moving point P for N masses system. Point P acceleration $\mathbf{r}_{p,tt}$ and vector $\mathbf{r}_{cg,P}$ are shown is appendix A. The total angular momentum is given by the following as follows:

$$\mathbf{H_P} = \sum_{i=1}^{N} m_i \mathbf{r_{ip}} \times \mathbf{r_{ip,t}} \tag{22}$$

Following equations(21)-(22), the total bending moment applied on point P is as follows:

$$\mathbf{M_p} = \left( M_{tot} y_{cg} \left( x_{tt} + u_{g,tt} \right) - m_1 h_1 v_{tt} + m_2 h_2 w_{tt} \right)(-\mathbf{e_3}) \tag{23}$$

As on can see, the first term in equation(23) represents inertial elastic forces applied by the tank foundations on the tank base, whereas the other terms represent bending moment exerted due to horizontal internal forces of sloshing and TMD vibration, respectively. Hence, for moment of inertia $I = \pi R^3 t$, the corresponding normal stress applied in point P due to total bending moment $\mathbf{M_p}$ is as follows:

$$\sigma_b = \frac{|\mathbf{M_p}|R}{I} = \frac{M\Omega^2}{\pi t_w}\left(\frac{h_c}{R}\right)\left|\left(1 + \varepsilon_1 \frac{h_1}{h_c} + \varepsilon_2 \frac{h_2}{h_c}\right)(\bar{x}'' + \bar{u}_g'') - \varepsilon_1 \frac{h_1}{h_c} v'' + \varepsilon_2 \frac{h_2}{h_c} w''\right| \tag{24}$$

Additional normal stress $\sigma_W$ is applied on the tank foundations due to system overall weight. For base area of $A_{base} = \pi R^2$ this stress is given as follows:

$$\sigma_W = \frac{M_{tot} g}{A_{base}} = \frac{(M + m_1 + m_2)g}{\pi R^2} \tag{25}$$

Since in the discussed case, body forces are negligible with respect to inertial stresses, the total normal stress applied in point P is as follows:

$$\sigma_N \approx \sigma_b \tag{26}$$

The equivalent Von-Mises stress $\sigma_{eq} = \sqrt{\sigma_N^2 + 3\tau_s^2}$ in point P is given after second time normalization according to $\tau = t_N / \sqrt{1+\varepsilon_1}$ as follows:

$$\sigma_{eq} = \frac{M\Omega^2}{\pi t_w (1+\varepsilon_1)} \times$$
$$\times \sqrt{\left(\frac{h_c}{R}\right)^2 \left(\left(1 + \varepsilon_1 \frac{h_1}{h_c} + \varepsilon_2 \frac{h_2}{h_c}\right)(\ddot{\bar{x}} + \ddot{\bar{u}}_g) - \varepsilon_1 \frac{h_1}{h_c} \ddot{v} + \varepsilon_2 \frac{h_2}{h_c} \ddot{w}\right)^2 + \frac{3}{4}(\ddot{\bar{x}} + \ddot{\bar{u}}_g)^2} \tag{27}$$

The PEA installation height can be chosen by the designer according to engineering constraints. From practical reasons and the will to reduce bending moments applied on the tank foundations due to PEA motion, we deside to locate the PEA near the tank base, $h_2 = 0$.

Ratio $\eta$ is defined between the controlled and uncontrolled system equivalent Von-Mises stress is as follows:

$$\eta(\tau) = \frac{\sigma_{eq,controlled}}{\sigma_{eq,uncontrolled}} \qquad (28)$$

## 2.3. Conditions for internal resonances

Natural frequencies of the system $\omega_i$ are calculated for the TMD attachment case ($\kappa_2 = 0$) by: $\det(\mathbf{K} - \omega^2 \mathbf{I}) = 0$, and calculated numerically for several values of parameters.

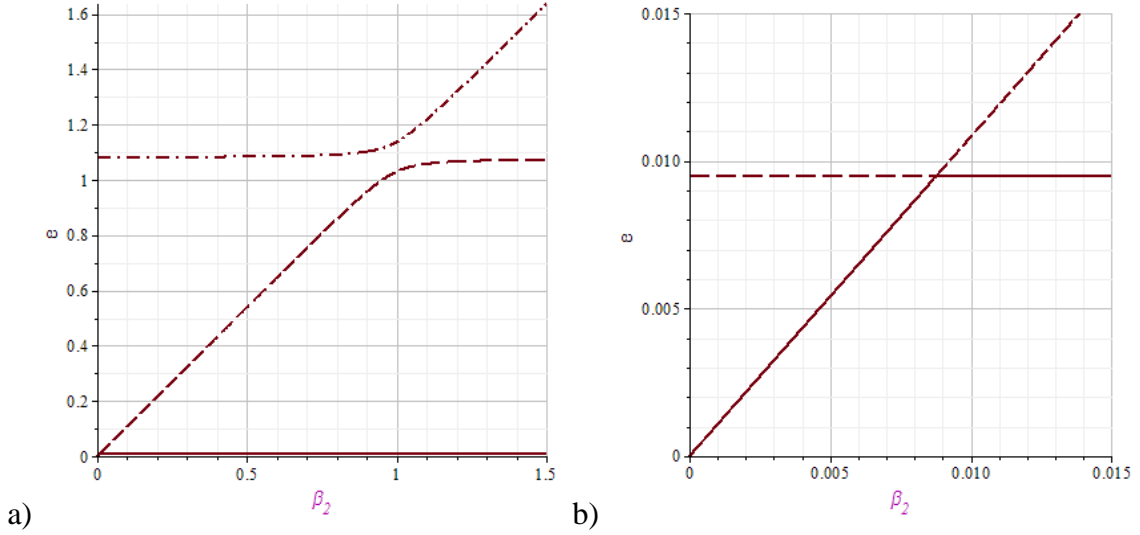

a) b)

*Figure 6- Natural frequencies vs. parameter $\beta_2$; $\omega_1$ (solid-line), $\omega_2$ (dashed-line) and $\omega_3$ (dashed-dotted line); a) for $\varepsilon_2 = 0.01$. For 75% full vessel and fundamental beam-type mode frequency of $\Omega = 1256.64 [rad/s]$ and $\beta_1 = 0.00875$. b) zoom-in for both cases for lower values of $\beta_2$.*

System (16) natural frequencies $\omega_i$ are defined to be an increasing series with index $i$. Hence, their ratios are presented in Figure 7 for filling volume percentage of 75%, fundamental beam-type vibration mode frequency of $\Omega = 1256.64 [rad/s]$ and TMD mass ratio of 1% from the structure and non-sloshing liquid mass:

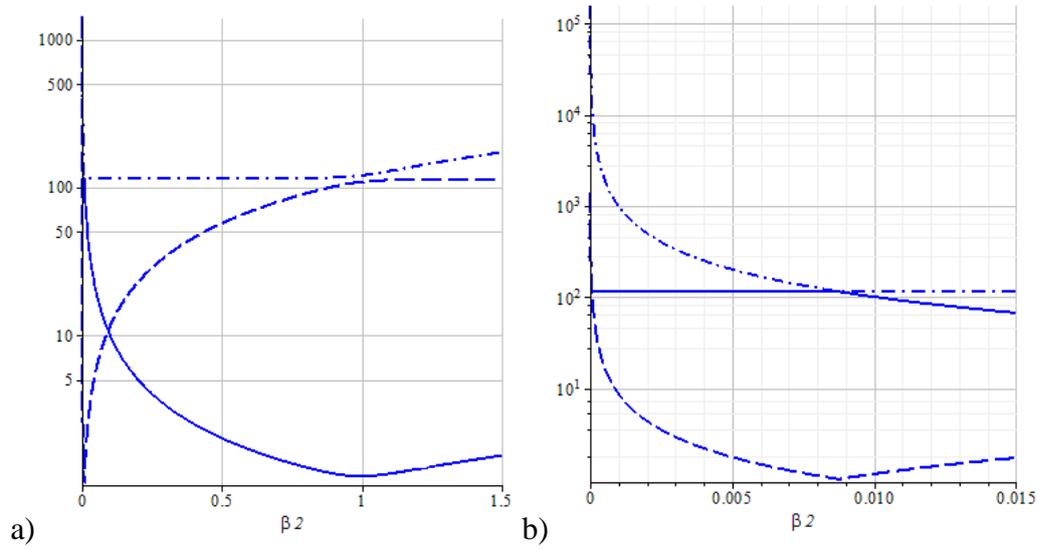

*Figure 7- Natural frequency ratios vs. parameter $\beta_2$ for $\varepsilon_2 = 0.01$, 75% full vessel and fundamental and $\beta_1 = 0.00875$; a) combined display of natural frequency ratios, dashed line: $\omega_2/\omega_1$, dashed-dotted line: $\omega_3/\omega_1$, solid-line: $\omega_3/\omega_2$, b) zoom-in for $\beta_2 = [0, 0.015]$*

# 3. Finite element modeling and analysis

In this section, finite element (FE) approach is used to determine the tank fundamental beam-type (1,1) mode natural frequency, for different values of filling percentage. The tank is considered to be clapped-free. Later on, the numerical results will be compared with approximated analytical results found by Chiba et al.[39]. Experimental and numerical investigation of modal properties for liquid contained structures for a several liquid filling heights was made by Jalali and Parvizi[40] and Mazuch[41]. The natural frequency values dependence in the filling percentage was investigated by Haroun(1980)[42] for both broad and tall tank. Analytical study and comparison to numerical findings for various shell vibration modes, filling percentages and boundary conditions was made by Housner and Haroun(1979-1981)[43,44], Haroun(1983) [45] and Goncalves and R. C. Batista(1985) [46]. Since there is no similitude between our experimental system and other vessel-liquid systems shown in literature, one cannot conclude exactly the modal characteristics of our particular system. However, between both the present system and the test system examined by Chiba et al. in reference[47] there is a similarity of boundary conditions and not bad similarity of tank geometrical and material properties. Hence, in the following, the dynamic response and modal analysis of a cylindrical shell partially filled with water is examined using FE method. FE analyses using the commercial computer code Multi Physics COMSOL 5.0 are performed to determine the structural modal modes and frequencies. The excitation was performed using pressure acoustic interaction applied through the air around the tank-liquid system. We consider the tank to be a perfectly anchored circular cylindrical thin elastic shell with a constant thickness. The shell material is assumed to be homogeneous, linearly elastic and isotropic. Four-node tetrahedral element is used to mesh both the shell and the contained liquid. The FE model of a half full cylindrical tank is shown if Figure 8.

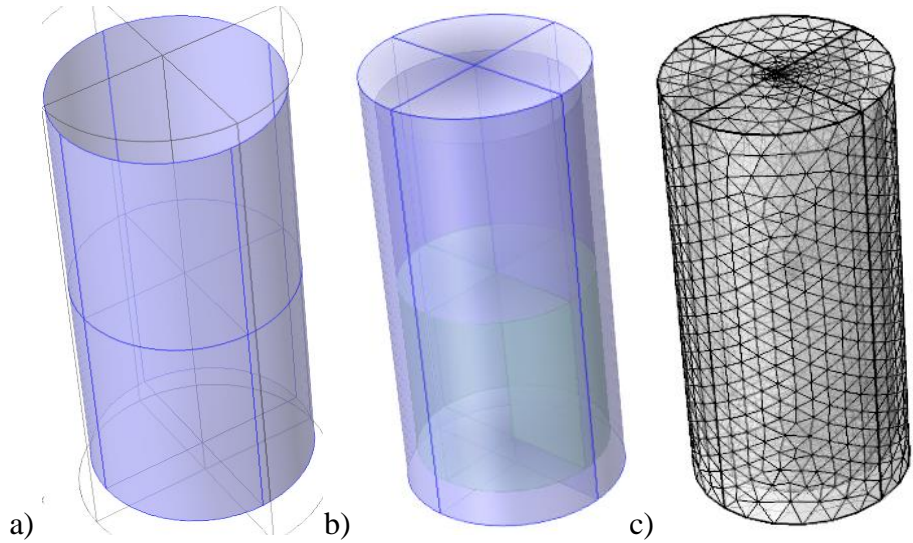

*Figure 8- The FE model of a half full cylindrical tank; a)Isometric view of the tank shell, b)the tank shell, filling liquid and the surrounding acoustic medium; c) Four-node tetrahedral mesh*

The material properties of the aluminum shell are as follows: Young's modulus of $E = 70 GPa$, Poisson ration of $\nu = 0.35$ and density of $\rho_{Al} = 2700\, kg/m^3$. The filling water density is $\rho_F = 997\, kg/m^3$. The natural frequency corresponding to the tank fundamental beam-type mode versus the filling percentage is shown in Figure 9.

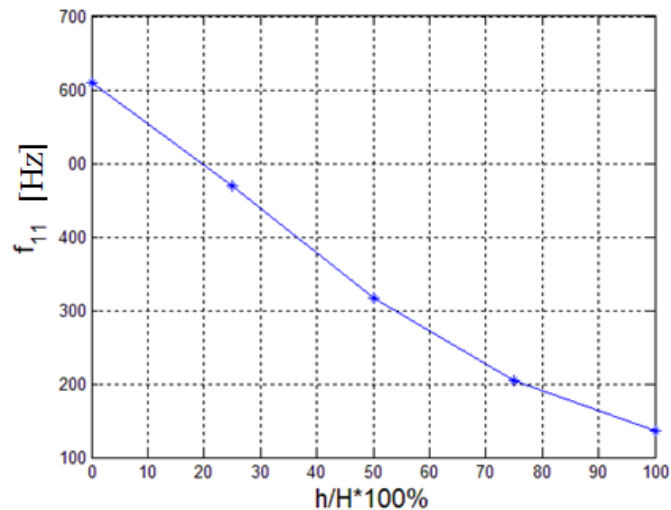

*Figure 9-Cylindrical partially-filled water tank fundamental beam-type (1,1) mode natural frequency versus filling percentage of the tank.*

Our experimental results were compared with the theoretical, numerical and experimental study made by M. Chiba, N. Yamaki and J. Tari [39,47,48] for similar test cylinders. Comparison between the current study experimental results and the literature analytical values is shown in Figure 10:

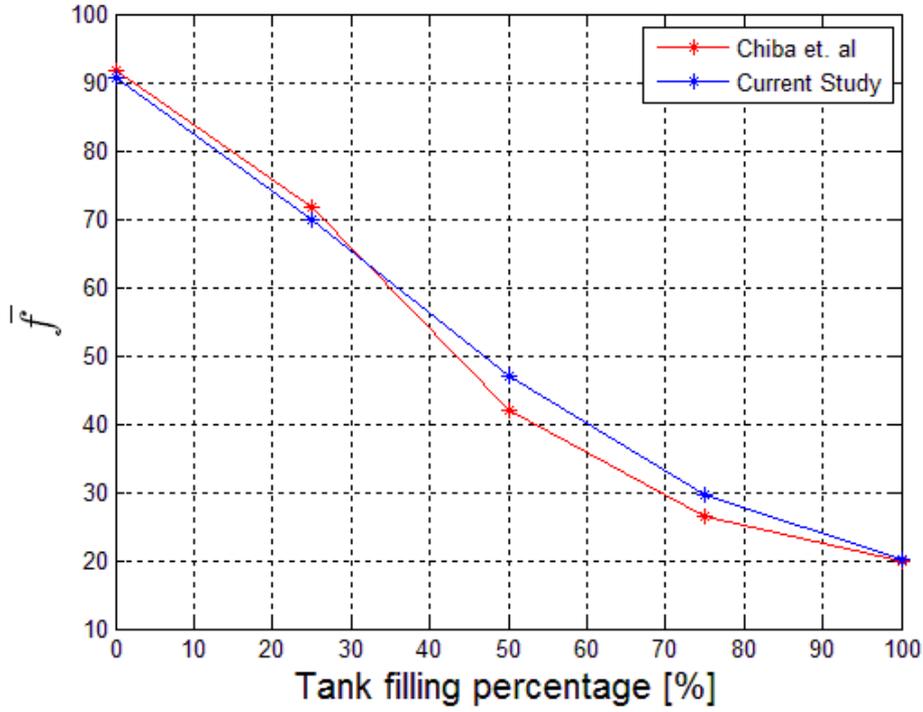

*Figure 10- Comparison between the non-dimensional frequency $\bar{f}$ found in current study and numerical results shown by Chiba et al.* [47].

One can see that near limiting cases of both the empty and the full case there is a good agreement between the theoretical values and the FE results. However, for other cases, there is a bias which increases as we get farther from those limiting cases. A source for this error is that the work done by Chiba et al. treated a test cylinders that differ from the present cylinder by their geometrical and material properties, represented by the non-dimensional parameter $\tilde{Z} = \frac{H^2}{Rt_w}\sqrt{1-\nu^2}$; while for the test cylinders presented by Chiba et al. $\tilde{Z} = 500$ and $\tilde{Z} = 2000$, for the present tank $\tilde{Z} = 2289.45$. Hence, a linear extrapolation performed. Frankly, comparison between both cases shown in literature, the sensitivity of the natural frequency with respect to parameter $\tilde{Z}$ is fairly small.

For farther numerical analysis, the tank-liquid combined system natural frequency of the relevant mode is required, i.e. (1,1) beam-type mode. One can easily understand that this value is strongly related to both foundation stiffness and tank filling percentage. In Figure 11 this relation is analyzed with the help of FE simulation, where the liquid-filled tank natural frequency is computed for different filling percentages versus tank foundation stiffness. One can see that for very large foundation stiffness values the tank-liquid system natural frequency reaches the values given for anchored tank as shown in Figure 10, since perfectly-fixed boundary condition is achieved.

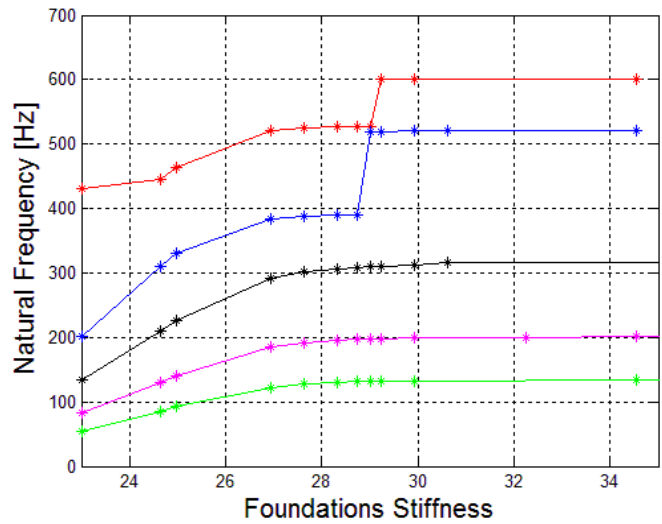

*Figure 11- FE simulation results of partially liquid-filled tank natural frequency in Hz versus logarithmic foundation stiffness in MPa, for different filling percentages (color online); empty tank (red), 25% full (blue), 50% full (black), 75% full (magenta), full tank (green).*

# 4. Numerical analysis

## 4.1. Structure and liquid sloshing parameter values

Further analysis will refer to an experimental system of a 75% water filled cylindrical storage tank. The tank has a diameter of $300mm$ and height of $H = 600mm$. It is made of an aluminum sheet $t_w = 1mm$ in thickness, with mass density of $\rho_{Al} = 2700\, kg/m^3$. For those dimensions and properties, the equivalent model parameters shown at the previous chapter where estimated formulas formulated by Dodge[11] and Abramson[12]. Following equation(3), for liquid height of $h = 450mm$ the total liquid mass is $m_T = 31.713kg$. The tank mass is as follows: $m_{tank} = 2\pi R H t \rho_{Al} = 1.53kg$. Following equations(2) and(5), the sloshing mass $m_1$ and the "static" liquid portion $m_0$ are calculated: $m_1 = 4.8kg$, $m_0 = 26.91kg$. The sloshing-mass height, following equation(6) is: $h_1 = 63.25mm$. Following equation(1), the total static mass of the tank-liquid system is: $M = 28.44kg$, and its height with respect to the tank foundations is calculated using equations (6) and (7): $h_c = 26.8mm$.

According equation(2), the fundamental asymmetric sloshing mode radial natural frequency can be estimated as follows:

$$\omega_1 = \sqrt{k_1/m_1} = \sqrt{\frac{1.849g}{R}\tanh(1.84h/R)} \qquad (29)$$

By substituting the tank dimensions to equation(29), we get that for 75% full tank the fundamental asymmetric sloshing mode natural frequency in units of $rad/s$ is: $11[rad/s]$ or $f_1 = \omega_1/2\pi = 1.75[Hz]$. According to the previous section, in order to calculate the non-dimensional system parameters one should have the fundamental beam-type modal frequency of the tank, corresponding to its (1,1) mode.

## 4.2. Non-dimensional Parameter values calculation

According to equation(12), the sloshing mass ratio is: $\varepsilon_1 = 0.17$. With respect to the fundamental beam-type mode natural frequency given from the FE analysis shown above we take $\Omega = 1256.64 \, [rad/s]$ corresponding to $200 \, [Hz]$ frequency given from the FE analysis, and consequently the following frequency ratio is obtained: $\beta_1 = 8.75 \cdot 10^{-3}$. Following Gendelman and Alloni[49] we take the impact terms coefficients as $\lambda = 20$ and $\kappa = 5$, and following Pilipchuk and Ibrahim [9] we take the impact smooth potential function power to be $n = 6$. Following Cho [50], the Rayleigh damping ratio is taken as $\bar{Z} = 2\%$ damping for the flexible-impulsive interaction modes and $\bar{\zeta}_1 = 0.5\%$ damping for the sloshing modes. Hence, the corresponding non-dimensional coefficients are taken as $Z = 3.7 \cdot 10^{-2}$ and $\zeta_1 = 5.4 \cdot 10^{-3}$.

## 4.3. Evaluation Criteria

In order to evaluate the energy portion absorbed by the PEA, we formulate the PEA and primary structure energy portion. We formulate the dimensional energies on the primary system and the PEA:

$$\tilde{E}_p = \frac{M}{2}(u_{g,t} + x_t)^2 + \frac{m_1}{2}(u_{g,t} + y_t)^2 + \frac{K}{2}x^2 + \frac{k_1}{2}(y-x)^2 + \frac{m_1 \tilde{k}}{2}\left(\frac{y-x}{R}\right)^{4n+2}$$

$$\tilde{E}_{PEA} = \frac{m_2}{2}(u_{g,t} + z_t)^2 + \frac{k_2}{2}(z-x)^2 + \frac{k_3}{4}(z-x)^4 \quad (30)$$

$$\tilde{E}_{tot} = \tilde{E}_p + \tilde{E}_{PEA}$$

when tilde stands for dimensional variable. Normalization and both of the time rescaling yields the following non-dimensional energy expression:

$$E_p = (\dot{x} + \dot{\bar{u}}_g)^2 + \varepsilon_1(\dot{\bar{y}} + \dot{\bar{u}}_g)^2 + (1+\varepsilon_1)\bar{x}^2 + \varepsilon_1(1+\varepsilon_1)\beta_1^2(\bar{y}-\bar{x})^2 + \frac{\varepsilon_1(1+\varepsilon_1)\bar{k}}{2n+1}(\bar{y}-\bar{x})^{4n+2}$$

$$E_{TMD} = \varepsilon_2(\dot{\bar{z}} + \dot{\bar{u}}_g)^2 + \varepsilon_2(1+\varepsilon_1)\beta_2^2(\bar{z}-\bar{x})^2 + \frac{1}{2}\varepsilon_2 \kappa_2 (\bar{z}-\bar{x})^4 \quad (31)$$

Where $E_p = \frac{2(1+\varepsilon_1)}{MR^2\Omega^2}\tilde{E}_p$, $E_{PEA} = \frac{2(1+\varepsilon_1)}{MR^2\Omega^2}\tilde{E}_{PEA}$. The PEA energy ratio with respect to the total energy in the system, noted by $\nu$, gives an insight about the PEA efficiency refers to the maximum amount of energy captured. It is the described by the following expression:

$$\nu(t_N) = \frac{\varepsilon_2(\dot{\bar{u}}_g + \dot{\bar{z}})^2 + \varepsilon_2(1+\varepsilon_1)\beta_2^2(\bar{z}-\bar{x})^2 + \frac{1}{2}\varepsilon_2\kappa_2(\bar{z}-\bar{x})^4}{\left\{\begin{array}{l}(\dot{\bar{x}}+\dot{\bar{u}}_g)^2 + \varepsilon_1(\dot{\bar{y}}+\dot{\bar{u}}_g)^2 + \varepsilon_2(\dot{\bar{u}}_g+\dot{\bar{z}})^2 + (1+\varepsilon_1)\bar{x}^2 + \varepsilon_1(1+\varepsilon_1)\beta_1^2(\bar{y}-\bar{x})^2 + \\ + \frac{\varepsilon_1\kappa}{2n+1}(\bar{y}-\bar{x})^{4n+2} + \varepsilon_2(1+\varepsilon_1)\beta_2^2(\bar{z}-\bar{x})^2 + \frac{1}{2}\varepsilon_2\kappa_2(\bar{z}-\bar{x})^4\end{array}\right\}} \quad (32)$$

Coordinate transformation (13) applied on expression yields:

$$\nu(\tau) = \frac{\varepsilon_2(\dot{\bar{u}}_g + (\dot{u}+\varepsilon_1\dot{v})/(1+\varepsilon_1) + \dot{w})^2 + \varepsilon_2(1+\varepsilon_1)\beta_2^2 w^2 + \frac{1}{2}\varepsilon_2\kappa_2 w^4}{\left\{\begin{array}{l}(\ddot{\bar{u}}_g + (\dot{u}+\varepsilon_1\dot{v})/(1+\varepsilon_1))^2 + \varepsilon_1(\ddot{\bar{u}}_g + (\dot{u}-\dot{v})/(1+\varepsilon_1))^2 + \varepsilon_2(\ddot{\bar{u}}_g + (\dot{u}+\varepsilon_1\dot{v})/(1+\varepsilon_1) + \dot{w})^2 + (u+\varepsilon_1 v)^2/(1+\varepsilon_1) + \\ +\varepsilon_1(1+\varepsilon_1)\beta_1^2 v^2 + \frac{\varepsilon_1\kappa}{2n+1}v^{4n+2} + \varepsilon_2(\ddot{\bar{u}}_g + (\dot{u}+\varepsilon_1\dot{v})/(1+\varepsilon_1) + \dot{w})^2 + \varepsilon_2(1+\varepsilon_1)\beta_2^2 w^2 + \frac{1}{2}\varepsilon_2\kappa_2 w^4\end{array}\right\}} \quad (33)$$

In impulsive forcing, for which the system gets a discreet energy amount in time zero, the energy dissipation rate is an important value for performance analysis. Variable $\Pi$ is defined to be the normalized energy held by the overall (vessel-liquid- PEA) system:

$$\Pi(\tau) = \tilde{E}_{tot}/\tilde{E}_{tot,0} = \frac{\left\{\begin{array}{l}\dot{\bar{x}}^2 + \varepsilon_1\dot{\bar{y}}^2 + \varepsilon_2\dot{\bar{z}}^2 + (1+\varepsilon_1)\bar{x}^2 + \varepsilon_1(1+\varepsilon_1)\beta_1^2(\bar{y}-\bar{x})^2 + \frac{\varepsilon_1\kappa}{2n+1}(\bar{y}-\bar{x})^{4n+2} + \\ +\varepsilon_2\dot{\bar{z}}^2 + \varepsilon_2(1+\varepsilon_1)\beta_2^2(\bar{z}-\bar{x})^2 + \frac{1}{2}\varepsilon_2\kappa_2(\bar{z}-\bar{x})^4\end{array}\right\}}{\left\{\begin{array}{l}\dot{\bar{x}}_0^2 + \varepsilon_1\dot{\bar{y}}_0^2 + \varepsilon_2\dot{\bar{z}}_0^2 + (1+\varepsilon_1)\bar{x}_0^2 + \varepsilon_1(1+\varepsilon_1)\beta_1^2(\bar{y}_0-\bar{x}_0)^2 + \frac{\varepsilon_1\kappa}{2n+1}(\bar{y}_0-\bar{x}_0)^{4n+2} + \\ +\varepsilon_2\dot{\bar{z}}_0^2 + \varepsilon_2(1+\varepsilon_1)\beta_2^2(\bar{z}_0-\bar{x}_0)^2 + \frac{1}{2}\varepsilon_2\kappa_2(\bar{z}_0-\bar{x}_0)^4\end{array}\right\}} \quad (34)$$

In order to optimize the PEA's parameters, we should have evaluation criteria [51–53], based on seismic responses of the combines system, the partially filled vessel and the PEA. Optimal parameter value will be achieved by minimizing an objective function which is sum of several evaluation criteria. These have to be suitable for our particular system.

Note that the numerator contains information from the PEA-coupled system's time history, when the denominator contains value recorded from the uncontrolled system.

PEA contribution to the overall system is roughly divided into dynamic and strength influence. The dynamical influence in manifested in lowering the time of energy dissipation or energy decay, and strength-manner is manifested in mitigation of the lateral reaction force applied on the tank wall.

The first evaluation criterion is the ratio between the controlled (PEA coupled) and uncontrolled system equivalent Von-Mises stress:

$$J_1 = \frac{\max_t |\sigma_{VonMizes}|_{with\ PEA}}{\max_t |\sigma_{VonMizes}|_{without\ PEA}} \quad (35)$$

This ratio represents the prevented stress rate. To assess the TET efficiency for the impulsive forcing case in more detailed manner, we define a characteristic time of the energy dissipation $\tau_c$ as follows:

$$\Pi(\tau_c) = \frac{1}{e} \quad (36)$$

where *e* is Euler number. This choice of the characteristic time might seem somewhat arbitrary; however, it is inspired by common under-damped linear oscillator. For this latter system, the inverse of characteristic time $\tau_c^{-1}$ is just equal to the coefficient of linear viscous damping. Critical dissipation time $\tau_c$ is defined as the period of time after-which the energy equivalently or sufficiently dissipated from the system. Hence, the second evaluation criterion is defined to be as follows:

$$J_2 = \frac{\tau_{c,\ with\ TMD}}{\tau_{c,\ without\ TMD}} \quad (37)$$

For the sake of optimization, according to the objective function is defined as follows, for a general case of N evaluation criteria: $\overline{OF} = \sum_{i=1}^{N} J_i$. Optimal PEA design for a specific excitation is the one that minimize $\overline{OF}$. However, since both evaluation criteria get values in different ranges, i.e. have smaller sensitivity to PEA efficiency, weight factors $w_i$ should be used: $\overline{OF} = \sum_{i=1}^{N} w_i J_i$, where $w_i$ is between zero to unity. For two evaluation criteria case, one can define the objective function as follows:

$\overline{OF} = w_1 J_1 + (1 - w_1) J_2$. The maximum value of $v$ is used to understand the PEA contribution to energy absorption efficiency.

## 4.4. Numerical Optimization for impulsive Excitation

Impact imposed on the vessel structure correspond to an impulse $F = A\delta(t)$, where $\delta(t)$ is Dirac's delta function. Following Vakakis et al. [38], this impulsive excitation can be modeled by non-zero initial velocity given to the structure: $\dot{x} = \tilde{\alpha}$. Hence, for system initially at rest, the initial conditions of the normalized and transformed coordinates are as follows: $u(0) = v(0) = w(0) = 0$ and $\dot{u}(0) = \dot{v}(0) = \alpha, \dot{w}(0) = -\alpha$.

Both evaluation criteria $J_1$ and $J_2$ were computed for various TMD and NES design parameters and impulse magnitudes $\alpha$. The numerical optimization conclusions for each energy absorber and evaluation criteria are given in the following sections. The optimization graphs for every impulsive magnitude $\alpha$ and energy absorber damping coefficient $\zeta_2$ (TMD) or $\xi_2$ (NES) present the evaluation criteria value versus mass ratio $\varepsilon_2$ and TMD frequency ratio $\beta_2$, or NES normalized stiffness $\kappa_2$.

In the numerical optimization graphs shown below, for every impulse magnitude value $\alpha$, there is an optimal parameter design set, described by the dark blue areas in the optimization 2D graphs; these are the optimal-mitigation zone (OMZ), for which the minimal evaluation criteria values are obtained, and consequently to optimal PEA performance. As one can understand, as the OMZ size is larger- the PEA performance dependence on design imperfections becomes minor. Optimal PEA design will lead to minimal evaluation criteria values and even more important, will be consistent for different impulsive magnitude values, for reliable passive PEA system.

### 4.4.1. TMD Optimization

The evaluation criteria are compared in the following figure. For each $(\alpha, \zeta_2)$ combination the objective function was computed for parameter range of $(\varepsilon_2, \beta_2) \in (0-0.5, 0-8)$.

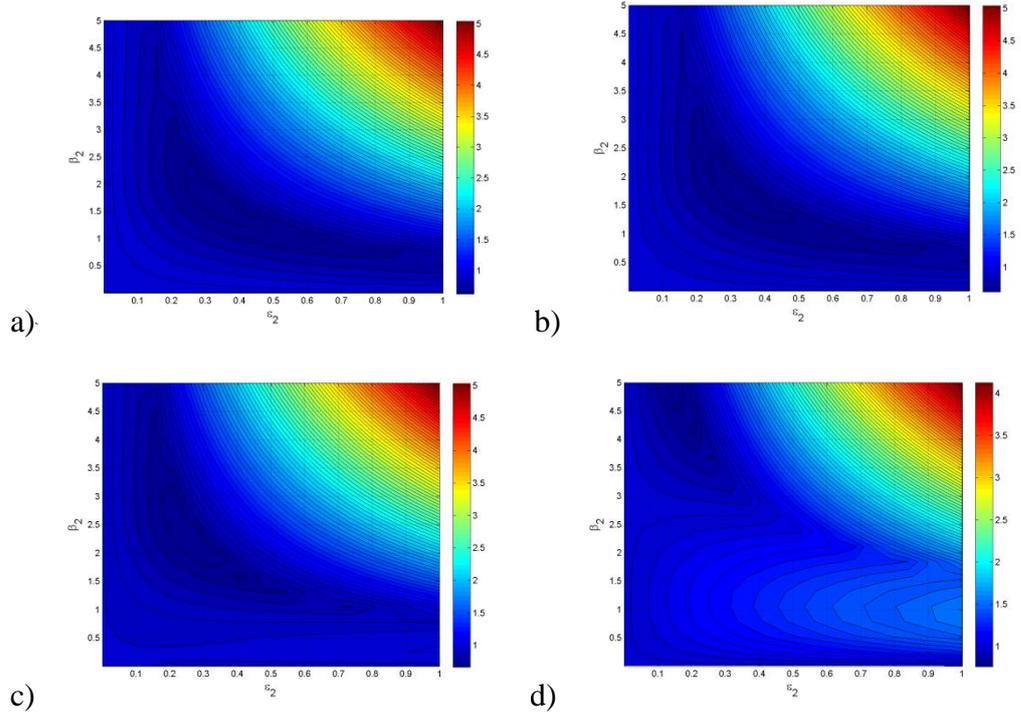

*Figure 12- Comparison table- $J_1$ evaluation criterion TMD graphs for $\zeta_2 = 0.75$ vs. $\varepsilon_2$ and $\beta_2$ within the [0,1] and [0,5], respectively; a) $\alpha = 0.1$, b) $\alpha = 0.5$, c) $\alpha = 1$, d) $\alpha = 2$.*

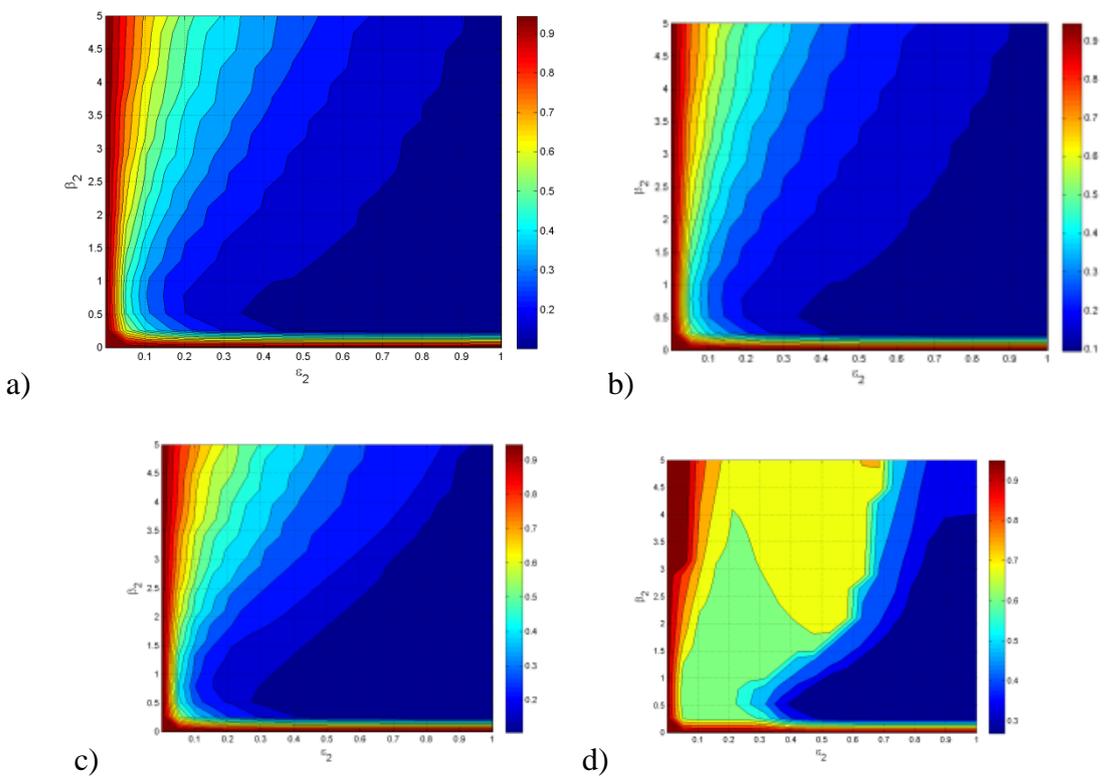

*Figure 13- Comparison table- $J_2$ evaluation criterion TMD graphs for $\zeta_2 = 1$ vs. $\varepsilon_2$ and $\beta_2$ within the ranges [0,1] and [0,5], respectively; a) $\alpha = 0.1$, b) $\alpha = 0.5$, c) $\alpha = 1$, d) $\alpha = 2$.*

Comparison between different efficiency performance graphs for various parameter and impulse amplitude values was made. It was clear that there is a trade-off between $J_1$ and $J_2$ optimizations. As a result, one can conclude that $w_1$ selection has a significant effect on the final TMD optimal design. Consequently, weight coefficient $w_1$ should be selected by the designer with respect to the engineering constraints, limitations and regulations or the structure functionality.

Regarding TMD optimization according to evaluation criterion $J_1$, From numerous numerical optimization graphs we observed that for every impulse magnitude value $\alpha$, there is an optimal parameter design set. Various numerical simulations show that damping coefficient value of $\zeta_2 = 1.25$ shows the optimal results, regarding both the minimal $J_1$ value achieved and the OMZ size.

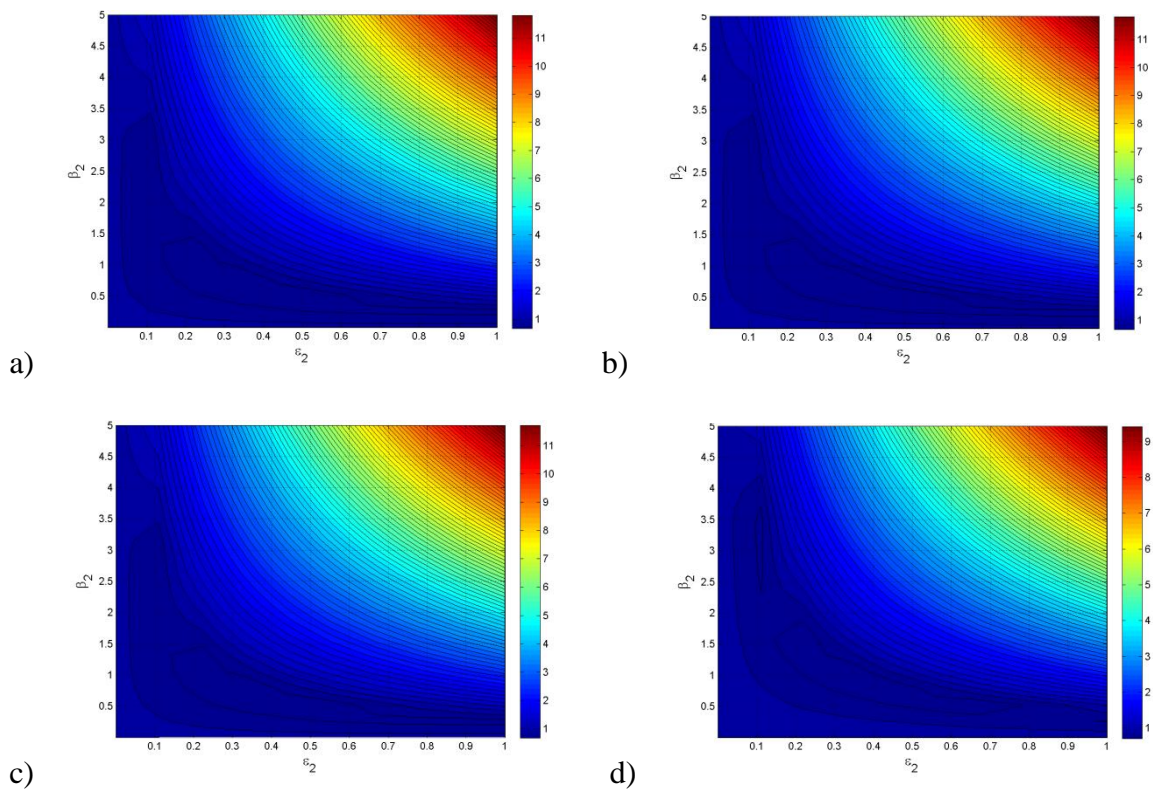

*Figure 14- $J_1$ evaluation criterion TMD optimization graphs for $\zeta_2 = 1.25$ vs. $\varepsilon_2$ and $\beta_2$ within the ranges [0,1] and [0,5], respectively; a) $\alpha = 0.1$, b) $\alpha = 0.5$, c) $\alpha = 1$, d) $\alpha = 2$.*

By observing Figure 14, one can see that for $\zeta_2 = 1.25$ there is a good OMZ shape and size stability. Hence, TMD design parameter values of $\varepsilon_2 = 0.35, \beta_2 = 0.78$ will be selected.

Regarding TMD optimization according to evaluation criterion $J_2$. From various numerical simulations we see that damping coefficient $\zeta_2$ values below 0.2 lead to poor mitigation performances. For specific impulsive magnitude $\alpha$ and different damping values, similar minimum values of $J_2$ are obtained. Hence, we select the optimal design parameter set according to the maximal OMZ size. For every $\zeta_2$ lager than unity we get the same OMZ size.

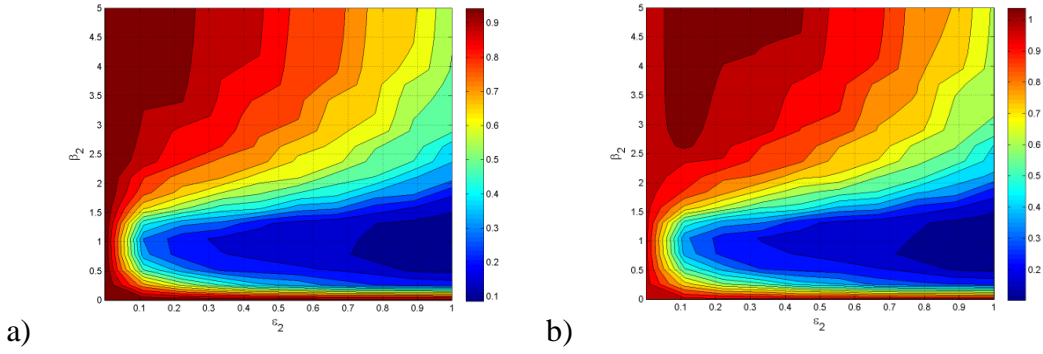

*Figure 15- $J_2$ evaluation criterion TMD optimization graphs for $\zeta_2 = 0.25$ vs. $\varepsilon_2$ and $\beta_2$ within the ranges [0,1] and [0,5], respectively; a) $\alpha = 0.5$, b) $\alpha = 1$.*

From Figure 15(a) one can see that for $\alpha$ value smaller than 0.5, and every damping coefficient $\zeta_2$, as we select larger $\varepsilon_2$ value- the time to decay will necessarily be lower. This phenomenon vanishes for larger impulse magnitude values Figure 15(b).

One can see that there is a knee, describes best mitigation performances for minimal TMD weight $\varepsilon_2$.

Observing optimization simulation for damping coefficients equal/lager than unity are fairly identical. Hence, optimal parameter set will be selected according to performance robustness with respect to impulse magnitude.

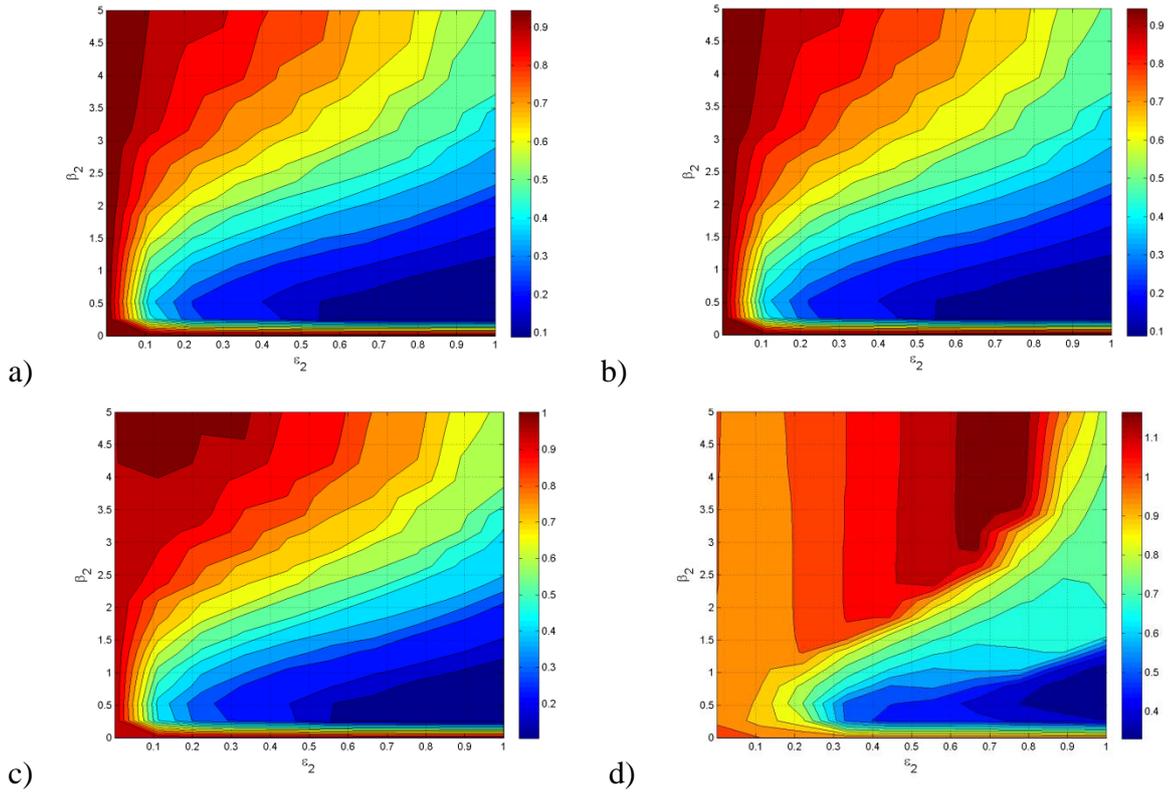

*Figure 16- $J_2$ evaluation criterion TMD optimization graphs for $\zeta_2 = 1.25$ vs. $\varepsilon_2$ and $\beta_2$ within the ranges [0,1] and [0,5], respectively; a) $\alpha = 0.1$, b) $\alpha = 0.5$, c) $\alpha = 1$, d) $\alpha = 2$.*

We select $\varepsilon_2 = 0.35, \beta_2 = 0.78$ to minimize both $J_1$ and $J_2$.

### 4.4.2. NES Optimization

Regarding NES optimization according to evaluation criterion $J_1$, for low magnitude impulsive excitation, as the TMD reviles minor dependence on stiffness coefficient $\kappa_2$. Moreover, as the damping coefficient $\xi_2$ becomes larger the OMZ corresponds to smaller mass ratios $\varepsilon_2$ (Figure 17). This independence on $\kappa_2$ leads to simplicity of the NES optimization graphs of $J_1$ with respect to those of the TMD. Moreover, there is a single optimal mass ratio.

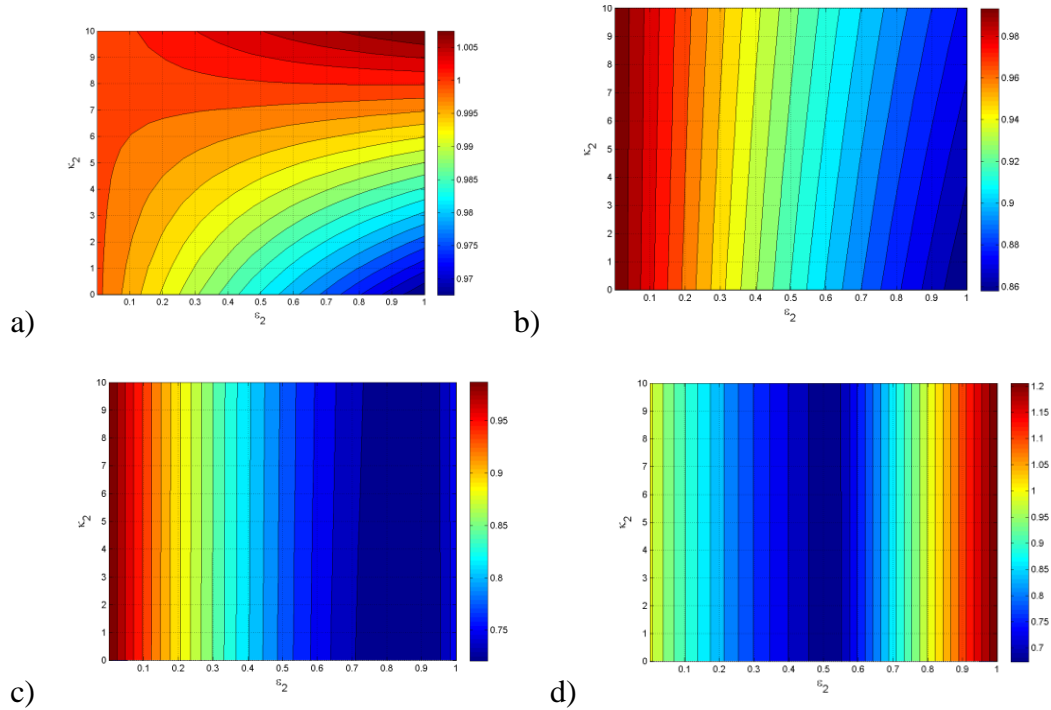

*Figure 17- $J_1$ evaluation criterion NES optimization graphs for small impulsive excitation magnitude $\alpha = 0.1$ vs. $\varepsilon_2$ and $\kappa_2$ within the ranges [0,1] and [0,10], respectively; a) $\zeta_2 = 0.05$, b) $\zeta_2 = 0.25$, c) $\zeta_2 = 0.75$, d) $\zeta_2 = 1.25$.*

As the impulse becomes larger, the OMZ begins to fold and shrink and shift to higher values of $\varepsilon_2$, mainly for smaller lower damping TMD designs (Figure 18).

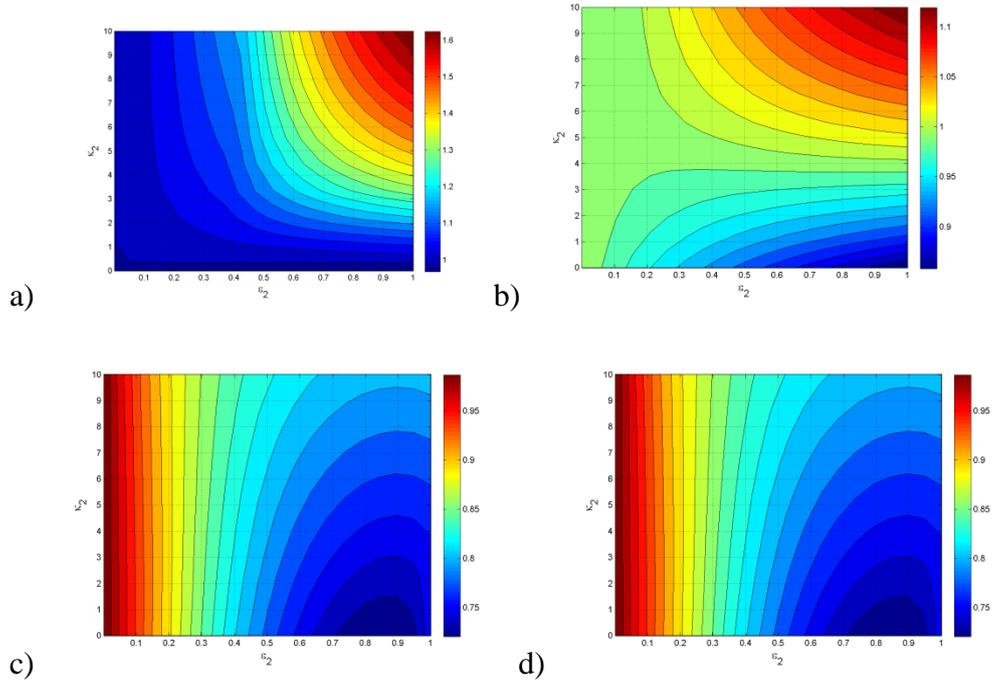

*Figure 18- $J_1$ evaluation criterion NES optimization graphs for intermediate impulsive excitation magnitude $\alpha = 0.5$ vs. $\varepsilon_2$ and $\kappa_2$ within the ranges [0,1] and [0,10], respectively; a) $\zeta_2 = 0.05$, b) $\zeta_2 = 0.25$, c) $\zeta_2 = 0.75$, d) $\zeta_2 = 1.25$.*

From comparing Figure 17 and Figure 18 we see that better OMZ values are obtained for higher values of parameter $\xi_2$. For all reasons mentioned abode, large damping coefficient TMD should be selected, i.e. $\varepsilon_2 = 0.4, \xi_2 = 1.75$ (Figure 19).

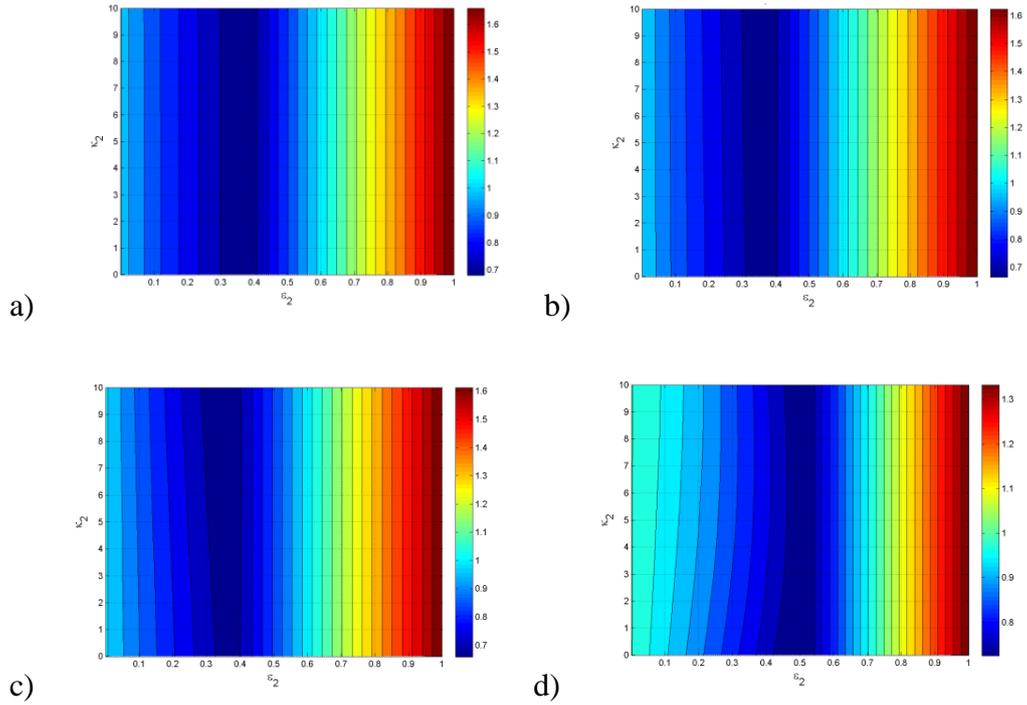

*Figure 19- $J_1$ evaluation criterion NES optimization graphs for $\zeta_2 = 1.75$ vs. $\varepsilon_2$ and $\beta_2$ within the ranges [0,1] and [0,10], respectively; a) $\alpha = 0.1$, b) $\alpha = 0.5$, c) $\alpha = 1$, d) $\alpha = 2$.*

Regarding NES optimization according to evaluation criterion $J_2$, for small values of $\alpha$, we see that as $\xi_2$ is bigger, the knee shape shows in the optimization graphs vanishes, in other words, the dependence on $\kappa_2$ becomes minor. Besides this fact, the NES OMZ values are identical for all $\xi_2$ selection.

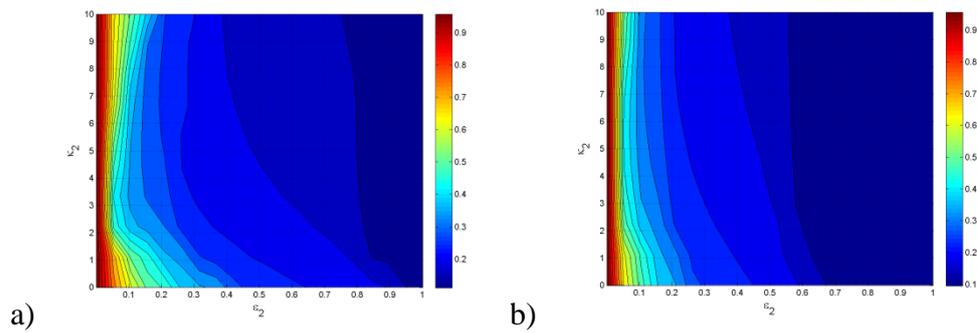

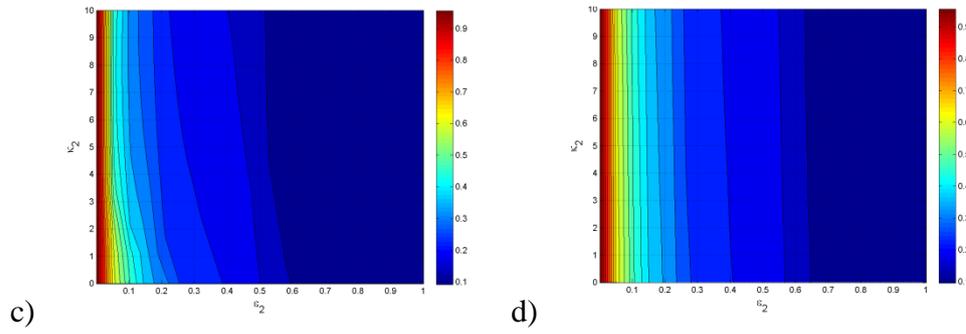

*Figure 20- $J_2$ evaluation criterion NES optimization graphs for small impulsive excitation magnitude $\alpha = 0.5$ vs. $\varepsilon_2$ and $\kappa_2$ within the ranges [0,1] and [0,10], respectively; a) $\xi_2 = 0.25$, b) $\xi_2 = 0.5$, c) $\xi_2 = 0.75$, d) $\xi_2 = 1.75$.*

This knee effect becomes more dominant for bigger $\alpha$ values. As a result, rather large damping coefficient should be selected. Comparing optimization graphs leads that the best OMZ values achieved for $\xi_2 = 1.25$ (Figure 21).

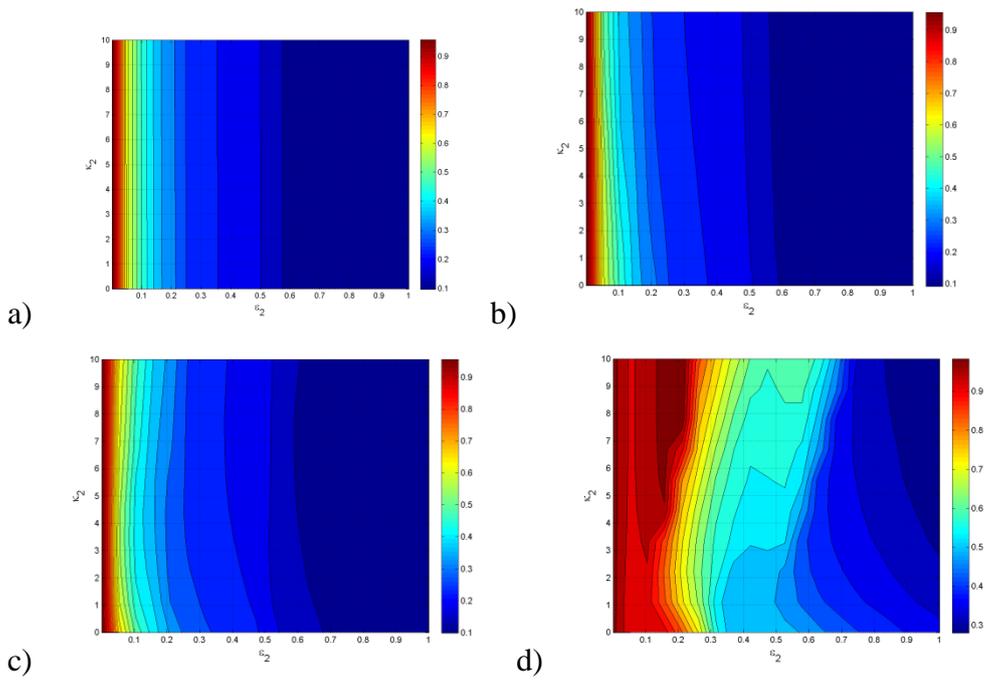

*Figure 21 – $J_2$ evaluation criterion NES optimization graphs for $\xi_2 = 1.25$ vs. $\varepsilon_2$ and $\beta_2$ within the ranges [0,1] and [0,10], respectively; a) $\alpha = 0.1$, b) $\alpha = 0.5$, c) $\alpha = 1$, d) $\alpha = 2$.*

## 5. Concluding remarks

The model developed in the paper aims to mimic both linear sloshing oscillations and strongly non-linear vibro-impact processes. This is achieved due to tailoring of the traditional linear mass-spring-dashpot model with the vibro-impact model, based on high-power smooth functions. We test both the TMD and NES attached to the primary tank-liquid system, as possible energy absorbers Careful selection of the damping coefficients ($\zeta_2$ for TMD and $\xi_2$ for NES) was found to have significant influence on the OMZ mitigation values and shape stability with respect to the non-dimensional impulse magnitude $\alpha$. It was demonstrated, that both the TMD and NES can oprovide efficient mitigation . In OMZ values, both TMD and NES provided similar results (mitigation of up to 40% with respect to $J_1$ and 90% with respect to $J_2$). However, the conservative TMD which is simple to design, suffers from significant sensitivity to frequency ratio selection $\beta_2$. On he other hand, the NES, which is more innovative and complicated to design, demonstrates minor dependence on the coupling coefficient $\kappa_2$, and thus offers better robustness.

## Acknowledgments

The authors are very grateful to Israel Science Foundation (Grant838/13) for financial support of this work. M. Farid is very grateful to Mr. Ofer Katsir for helpful discussions.

# Appendix A

$$\begin{aligned}
\mathbf{r}_{c,P} &= R\,\mathbf{e_1} + h_{c.g.}\mathbf{e_2} \\
\mathbf{r}_{1,P} &= (R-v)\mathbf{e_1} + h_1\mathbf{e_2} \\
\mathbf{r}_{2,P} &= (R+w)\mathbf{e_1} + h_2\mathbf{e_2}
\end{aligned} \tag{A1}$$

$$F_0 = -Kx - Cx_t$$

$$F_1 = \frac{m_1\tilde{k}(2n+1)}{R}\left(\frac{y-x}{R}\right)^{4n+1} k_1(y-x) + c_1(y_t - x_t) + m_1\tilde{c}(2n+1)(y_t - x_t)\left(\frac{y-x}{R}\right)^{2n} \tag{A2}$$

$$F_2 = k_2(z-x) + c_2(z_t - x_t)$$

$$\begin{aligned}
x_{cg} &= x + u_g + \frac{m_2 w - m_1 v}{M + m_1 + m_2} \\
y_{cg} &= \frac{M h_c + m_1 h_1 + m_2 h_2}{M + m_1 + m_2}
\end{aligned} \tag{A3}$$

$$\begin{aligned}
\ddot{\mathbf{r}}_p &= (\ddot{x} + \ddot{u}_g)\mathbf{e_1} \\
\mathbf{r}_{cg,P} &= R\,\mathbf{e_1} + h_{cg}\mathbf{e_2}
\end{aligned} \tag{A4}$$